\documentclass[printer]{aa}

\usepackage{graphicx}
\usepackage{txfonts}
\usepackage{natbib}
\usepackage{rotating}

\newcommand{\vinf}{\mbox{$v_{\infty}$}}

\usepackage{color}

\begin{document}

\authorrunning{Ignace, St-Louis, Proulx-Giraldeau}
\titlerunning{Polarimetric modeling of CIRs in winds}

\title{Polarimetric modeling of corotating interaction regions (CIRs)
threading massive-star winds}

\author{
   Richard Ignace\inst{1}
   \and
   Nicole St-Louis\inst{2}
   \and
   F\'{e}lix Proulx-Giraldeau\inst{2}
}

\institute{
   Department of Physics \& Astronomy, East Tennessee State University,
   Johnson City, TN 37614, USA
   \and
   D\'{e}partement de Physique, Pavillon Roger Gaudry, Universit\'{e} 
   Montr\'{e}al, C.P.\ 6128, Succ.\ Centre-Ville, Montr\'{e}al,
   Quebec, H3C 3J7
}

\date{}

\abstract
{
Massive star winds are complex radiation-hydrodynamic (sometimes
magnetohydrodynamic) outflows that are propelled by their enormously
strong luminosities.  The winds are often found to be structured and variable,
but can also display periodic or quasi-periodic behavior in a variety
of wind diagnostics.  
}
{
The regular variations observed in putatively 
single stars, especially in UV wind lines, have often been attributed
to corotating interaction regions (CIRs) like those seen in the solar
wind.  We present light curves for variable polarization from winds
with CIR structures.  
}
{
We develop a model for a time-independent CIR based on a kinematical
description.  Assuming optically thin electron scattering, we explore
the range of polarimetric light curves that result as the curvature,
latitude, and number of CIRs are varied.
}
{
We find that a diverse array of variable polarizations result from an
exploration of cases.  The net polarization from an unresolved source
is weighted more toward the inner radii of the wind.  Given that most
massive stars have relatively fast winds compared to their rotation
speeds, CIRs tend to be conical at inner radii, transitioning to a spiral
shape at a few to several stellar radii in the wind.
}
{
Winds with a single CIR structure lead to easily identifiable polarization
signatures.  By contrast allowing for multiple CIRs, all emerging from
a range of azimuth and latitude positions at the star, can yield complex
polarimetric behavior.  Although our model is based on some simplifying
assumptions, it produces qualitative behavior that we expect to be robust,
and this has allowed us to explore a wide range of CIR configurations that
will prove useful for interpreting polarimetric data.
}


\maketitle

\section{Introduction}

A corotating interaction region (CIR) arises from the interaction
of wind flows that have different speeds: rotation of the star
ultimately leads to a collision between the different speed flows
to produce a spiral pattern in the wind.  CIRs have been well-studied
in the solar wind
\citep[e.g.,][]{2008GeoRL..3510110R,1983GeoRL..10..413B,1997JGR...102.4661B,
1999SSRv...89...21G}.  They have also been invoked to explain a
variety of phenomena observed in massive star winds.  Evidence of large-scale structure in the winds of massive stars
is manifold.  For example, the ubiquitousness of discrete
absorption components (DACs) in the ultraviolet (UV) spectra of OB
stars \citep{1989ApJS...69..527H,1997A&A...327..699F}, some shown
to have a recurrence timescale related to the rotation rate
\citep{1999A&A...344..231K}, are thought to be a consequence of the
presence in the wind of CIRs.  \citet{1996ApJ...462..469C} predicted
that such regions should develop following a perturbation, such as
a bright or dark spot, on the stellar photosphere that propagates
in the wind, thus generating a complex spiral-like structure of low and
high density and velocity regions as the star rotates. CIRs have
also been associated with the periodic spectroscopic, photometric,
and polarimetric variability observed in some Wolf-Rayet (WR) stars
\citep{1995ApJ...452L..57S, 1997ApJ...482..470M, 1999ApJ...518..428M,
2010ApJ...716..929C}.

Nevertheless, the origin of the perturbations generating the departure
from spherical symmetry in massive-star winds remains unclear. Possible
physical processes include non-radial pulsations and magnetic fields. The
latter is particularly promising. Large-scale, organized magnetic fields
have been found in only 7~\%\ of O stars \citep{2013arXiv1310.3965W}, and
these are thought to have a fossil origin. However, it has been suggested
that smaller scale, localized magnetic structures could be responsible
for the onset of DACs \citep{1994Ap&SS.221..115K}.  More recently,
\citet{2011A&A...534A.140C} have presented a model that suggests that such
structures could be widespread for stars that harbor a subsurface
convection zone. Such zones have been predicted to exist in hot massive
stars by \citet{2009A&A...499..279C} owing to an opacity peak caused
by the partial ionization of iron-group elements.  Detecting such
small-scale magnetic structures at the surface of massive stars is an
extremely difficult task  when dealing with O stars \citep{2013msao.confE..77K} and
even more difficult for WR stars where the photosphere is hidden by
the optically thick wind.

Studying the wind structures themselves can provide useful constraints
that will shed new light on their
origin. Polarimatry is a powerful means of studying asymmetries in stellar
winds. Indeed, in view of their high temperatures, massive-star outflows
contain a copious number of free electrons that scatter light, thereby
generating linear polarization.  For unresolved and spherically symmetric
envelopes, there is total cancellation of the polarization as integrated
across the source.  Consequently, a net polarization intrinsic to the
source demonstrates deviation by the source geometry from the spherical.
\citet{1977A&A....57..141B} showed that for optically thin scattering and an
axisymmetric envelope, the net polarization depends on three factors\footnote{
The analysis of \citet{1977A&A....57..141B} does not include limb
darkening.  \citet{1989ApJ...344..341B} considered limb darkening, which can introduce chromatic effects,
since it is $\lambda$-dependent.}:
the viewing inclination, an average optical depth for the scattering
envelope, and the shape of the envelope.

Since electron scattering is gray, broadband continuum polarization
measurements can be used to characterize the asymmetry at any given
time. If time-dependent measurements are available, more accurate
constraints can be obtained because the structure will then be observed from
different viewing angles. Spectropolarimetry can also be used to study the
structure in the line-forming region and to obtain kinematic information. In
the case of CIRs, this is particularly important because they are thought
to originate in the photosphere and propagate throughout the line-forming
region of the wind.

To understand the phenomena of CIR and use their occurrence
as a diagnostic of wind and atmospheric structure, models are needed
to interpret observations.  \cite{1986A&A...165..157M}
was an early proponent of
CIRs observed in the solar wind as a viable explanation for certain
types of variability (noted above) observed in massive star winds.
An important breakthrough in the field came with 2D hydrodynamic
simulations by \citet{1996ApJ...462..469C}, who modeled an equatorial CIR
for a steady-state, line-driven wind.  Following on this, 
\citet{2004A&A...423..693D}
developed 3D simulations for CIRs off the equator.  Both of those
works explore the consequences for wind line-profile effects, such as DACs
and other periodic signatures.  \citet{2004A&A...413..959B} considered inverse
techniques for extracting kinematic information about CIRs from wind
emission lines.

Pertinent to this contribution, there has also been work on the
polarization that can arise from CIR structures.  \citet{2000MNRAS.315..722H}
explored
a number of effects that give rise to variable polarization for rotating
stars, including a model for an equatorial CIR structure.  Harries
explored these effects using Monte Carlo radiative transfer simulations.
In contrast, \citet{2009AJ....137.3339I}
adopted a similar kinematic prescription
for the structure of an equatorial CIR in an application for the variable
linear polarization seen in the blue supergiant star HD~92207.  In this
paper we extend these initial results in a parameter study of variable
polarization from a wind with CIRs.  Neither of the preceding papers
allowed for multiple CIRs, and both were restricted to a CIR in the
equatorial plane.  Here results are presented that allow for multiple
CIRs and at arbitrary latitudes.

The structure of the paper is as follows.  Section~\ref{sec:model}
presents the base model for the CIR structure and the resulting variable
polarization, following but expanding on \cite{2009AJ....137.3339I}. The
results of our parameter study are presented in Section~\ref{sec:results}.
Concluding remarks are given in Section~\ref{sec:conc}, with particular
consideration of applications for observations.  Our results are presented
for a wind model that is set by a commonly used velocity structure.
An Appendix provides a solution for a different velocity structure as
a comparison example.

\section{Polarization from a simple CIR}
\label{sec:model}

In the next section, the approach of \citet{1977A&A....57..141B}, modified
by \citet{1987ApJ...317..290C}, for optically thin electron scattering 
(i.e., single scattering) in
an axisymmetric geometry is reviewed.  Then, an application is considered for
a cone structure in an otherwise spherical wind.  Then we
show how to use the cone as a basis for computing variable polarization
from an equatorial CIR for a rotating star, and finally the polarization
for CIRs is explored at arbitrary latitudes.

\subsection{Polarization from optically thin electron scattering}

For optically thin scattering, the amount of scattered light from
a circumstellar envelope is determined by integrating the
volume emissivity over the extent of the envelope.
The volume emissivity $j$ for electron scattering depends on the electron
number density $n_{\rm e}$, the Thomson scattering cross-section
$\sigma_T$, and the amount of incident radiation, as well as geometrical
factors relating to dipole scattering toward the observer.
It is assumed that the star is the sole source of illumination
for circumstellar electrons, meaning that the circumstellar envelope is
sufficiently optically thin so that the amount of scattered light is small
compared to the starlight.

We consider a coordinate system with the Earth along the $z$-axis.
The star has an axis $z_\ast$ that is its rotation axis.
The angle of inclination between these axes is signified by
$\hat{z}\cdot\hat{z}_\ast = \cos i_0$.  It is further assumed that
$y=y_\ast$, and therefore $\hat{x}\cdot\hat{x}_\ast = \cos i_0$ as well.
Following \citet{1977A&A....57..141B} for a point source of illumination,
the scattering angle between a ray of starlight and the observer is
$\chi$.  A scatterer is oriented about the observer's axis with azimuth
$\psi$ from the $x$-axis.

In the Stokes vector prescription, the polarization properties
of intensity are described in terms of I, Q, U, and V
\citep[e.g.,][]{1960ratr.book.....C}.  The latter is for circular
polarization, which will not be considered further.  Stokes I is a measure
of the total intensity.  Stokes-Q and U describe the  linear polarization
of the light.  For example, the relative linear polarization is

\begin{equation}
p = \frac{\sqrt{Q^2+U^2}}{I},
\end{equation}

\noindent 
and the polarization position angle is 

\begin{equation}
\tan 2\psi_{\rm p} = \frac{U}{Q}.
\end{equation}

Again, we assume that the total flux of light will be dominated by
direct starlight, meaning (a) that the amount of scattered starlight is
small and (b) that the amount of absorbed starlight by the circumstellar
envelope is miniscule.  Consequently, the emissivity for the Stokes-I
component is not a concern here.  The emissivities for Stokes-Q
and U are given by

\begin{equation}
j_Q = \left(\frac{\sigma_T}{4\pi r^2}\right)\,L_\nu\, n_{\rm e}\times 
        \frac{3}{4}\, \sin^2\chi\,\cos 2\psi,
\end{equation}

\noindent and

\begin{equation}
j_U = \left(\frac{\sigma_T}{4\pi r^2}\right)\,L_\nu\, n_{\rm e}\times 
        \frac{3}{4}\, \sin^2\chi\,\sin 2\psi.
\end{equation}

\noindent Here $r$ is the radius for a scattering electron from the
center of the star, and $L_\nu$ is the monochromatic stellar luminosity
at frequency $\nu$.  As previously noted, the angle $\chi$ and $\psi$
pertain to the geometry of the scattering.

\begin{figure}
\centering
\includegraphics[width=\hsize]{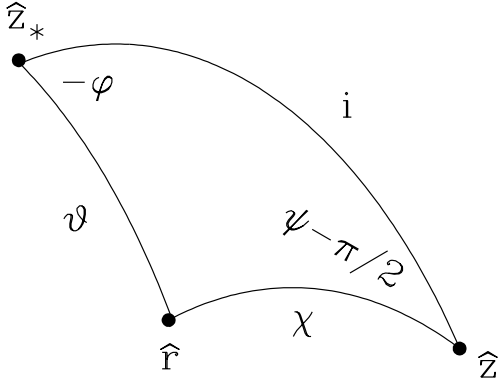}
\caption{Scattering geometry for Thomson scattering.
Here $\hat{z}_\ast$, $\hat{z}$, and $\hat{r}$ are unit vectors
for the stellar rotation axis, observer axis, and local
radial.  The scatterer has coordinates $(\vartheta,\varphi)$.
The scattering angles are $(\chi,\psi)$, and $i$ is the viewing
inclination.  The appearance of $\pi/2$ ensures that a net
polarization aligned with $\hat{z}_\ast$ has $q>0$, versus
$q<0$ if perpendicular.
}
\label{fig:geom}
\end{figure}

The star has spherical coordinates $(r,\vartheta, \varphi)$.
The observer is located at $\varphi=0^\circ$ and $\vartheta = i_0$
(see Fig.~\ref{fig:geom}).
Using $\mu=\cos\vartheta$, \citet{1977A&A....57..141B} showed that for
an axisymmetric distribution of electron scatterers, integration about
$\psi$ gives

\begin{equation}
\int_0^{2\pi} \, \sin^2\chi\,\cos 2\psi \, d\psi = -\pi \sin^2 i_0\, (1-3\mu^2),
\end{equation}

\noindent and

\begin{equation}
\int_0^{2\pi} \, \sin^2\chi\,\sin 2\psi \, d\psi = 0.
\end{equation}

\noindent These imply that only the U-polarization vanishes upon integration about the scattering volume
for an unresolved source, and only
the Q-polarization survives.  The total polarization $p$ is then given by

\begin{equation}
q = \frac{L_P}{L_\nu}=\frac{3}{16}\,\tau_0\,\sin^2 i_0\,
        \int\int\,\left[\frac{n_{\rm e}(\tilde{r},\mu)}{n_0}\right]\,(1-3\mu^2)\,d\tilde{r}\,d\mu,
        \label{eq:BM}
\end{equation}

\noindent where $L_Q$ is the luminosity of polarized light, $n_0$  some
reference number density of scatterers, $\tilde{r}=r/R_\ast$ 
the radius in the wind as normalized to the stellar radius, 
and $\tau_0$  an optical depth parameter.
The factor $\tau_0$ is given by

\begin{equation}
\tau_0 = n_0\,\sigma_T\,R_\ast.
\end{equation}

\citet{1987ApJ...317..290C} introduced a correction term to relax the
assumption of a point source star.  The correction is known as the
finite depolarization factor, $D(r)$.  The factor assumes a star of
uniform brightness (i.e., no limb darkening).  Although 
\citet{1977A&A....57..141B}
considered more general representations for this correction, we assume a star of uniform brightness.

The factor $D$ accounts for the varying degree of anisotropy of the
radiation field as perceived by a scattering electron at different
distances from the star.  
The factor is given by

\begin{equation}
D(\tilde{r}) = \sqrt{1-\frac{1}{\tilde{r}^2}}.
\end{equation}

\noindent This corrective modifies the polarization by being inserted as a
``weighting'' factor in the integrand of equation~(\ref{eq:BM}):

\begin{equation}
q = \frac{L_P}{L_\nu}=\frac{3}{16}\,\tau_0\,\sin^2 i_0\,
        \int\int\,\left(\frac{n_{\rm e}}{n_0}\right)\,D(\tilde{r})\,(1-3\mu^2)\,
        d\tilde{r}\,d\mu.\end{equation}

\noindent The net relative polarization is determined by $q$ with
$u=0,$ owing to the axisymmetry of the envelope.  We note that $q$ can
be positive or negative, signifying the orientation of the net
polarization against the sky.  The projected symmetry axis of the envelope is
taken lie along as the north-south direction
in the sky.  Consequently, a structure like
a bipolar jet would have $q<0$, whereas one like an equatorial disk
would have $q>0$.  The result shows that the polarization scales
in relation to viewing inclination as $\sin^2 i$.  Additionally, the
polarization is linear in the optical depth scaling parameter for the
envelope with $q \propto \tau_0$.  The ratio $q/\tau_0$ is thus
a function purely from the envelope geometry.

\subsection{The case of a simple cone}

Our treatment for CIRs is in terms of segments of a cone that are
phase-lagged from each other in a radially dependent way.  It therefore
is useful to first review the treatment of the polarization from a
simple cone.

We consider a cone with $z_\ast$ the symmetry axis of the cone: the cone
lies directly over the stellar rotation axis with opening angle
$\beta_0$.  The star has a spherically symmetric wind, and the cone
represents an axisymmetric sector of the wind that has either higher
or lower density, thereby breaking the spherical symmetry of the system
and leading to a net polarization from electron scattering of starlight
as described in the previous section.  In relation to the
results presented in that section, the main considerations 
here are (a) to relate the polarization to the properties of the
wind and (b) to use the result as a basis for our construction
of a CIR.

The stellar wind is taken to have mass-loss rate $\dot{M}$ and wind
terminal speed \vinf.  The gas is ionized
with a mean molecular weight per free electron $\mu_{\rm e}$.  The number
density of electrons in the spherical wind is given by

\begin{equation}
n_{\rm w} = \frac{\dot{M}/\mu_{\rm e}m_H}{4\pi\,R_\ast^2\vinf}\,\left[\frac{1}
        {\tilde{r}^2\,w(\tilde{r})}\right]\equiv n_0\,\tilde{r}^{-2}\,w^{-1},
\end{equation}

\noindent where $n_0$ is a parameter for the scale of the number density
that is given by the factors for wind and star parameters at the beginning
of the equation, 
and $w=v(\tilde{r})/\vinf$ is the
normalized wind velocity.  We use a standard wind velocity law given
by

\begin{equation}
w(\tilde{r}) = \left(1-\frac{b}{\tilde{r}}\right)^\gamma,
\end{equation}

\noindent with $\gamma \ge 0$ the velocity law exponent, and $b$ a
dimensionless parameter that sets the initial wind speed at the wind
base, with $w_0 = (1-b)^\gamma$.  Examples used in this paper
adopt $\gamma=1$ (although the Appendix presents a solution with
$\gamma=2$).

The electron density in the cone is parametrized by

\begin{equation}
n_{\rm c} = (1+\eta)\,n_{\rm w}(\tilde{r}),
\end{equation}

\noindent where $\eta$ is a dimensionless parameter taking values
in the interval $\eta \in [-1,\infty)$.  In effect, $\eta$ represents
an excess or decrement of density in the cone relative to the
otherwise spherical wind.  The cone has zero density if $\eta=-1$.
If $\eta=0$, the cone has the same density as the wind, in which
case sphericity is preserved, and one expects zero net polarization.
In this form we are taking the run of density with radius in the
cone to have the same functional form as in the wind.  This need
not be the case, and certainly a different radial dependence for
the density in the cone as compared to the wind could be treated
(e.g., perhaps $\gamma$ is different for the flow in the cone as
compared to the wind flow).

The results of Brown \& McLean and Cassinelli et al. are general
to any axisymmetric distribution of scatterers.  However, a cone
is an example of a special class of axisymmetric envelopes in which
the density is mathematically separable in terms of radial and
angular functions.  
With $\mu=\cos \theta$, we define the function $f$
and $g$ as

\begin{equation}
n_{\rm c}/n_0 = f(\tilde{r})\,g(\mu).
\end{equation}

\noindent In the application here, $f(\tilde{r}) = \tilde{r}^{-2} w^{-1}$ is the same
both in the cone and in the wind.  For a cone the function $g=1$
for $\mu \ge \cos \beta_0$ and $g=0$ for $\mu < \cos(\beta_0)$.

Consider now the polarization from the cone.  It is determined by a
volume integral over the entire envelope, in both the cone and
the non-cone regions.  The polarization is given by

\begin{equation}
q = \frac{3}{16}\,\tau_0\,\sin^2 i_0\,\int\int\,\frac{n_{\rm e}(\tilde{r},\mu)}
        {n_0}\,D(\tilde{r})\,(1-3\mu^2)\,d\mu\,d\tilde{r},
\end{equation}

\noindent where $n_{\rm e}$ is the electron density, which is either $n_{\rm
w}$ or $n_{\rm c}$.  Since the cone density scales with the wind density,
the integration can be recast as

\begin{eqnarray}
q & = & \frac{3}{16}\,\tau_0\,\sin^2 i_0\times \nonumber \\ 
  & & \left[ \int_1^\infty\int_{-1}^{+1}
        \,\frac{n_{\rm e}}{n_0}\,
        D(\tilde{r})\,(1-3\mu^2)\,d\mu\,d\tilde{r} \right. \nonumber\\ 
 & & \left. + \eta\,\int_1^\infty\int_{\mu_0}^{+1}
        \,\frac{n_{\rm e}}{n_0}\,
        D(\tilde{r})\,(1-3\mu^2)\,d\mu\,d\tilde{r}\right],
\end{eqnarray}

\noindent where $\mu_0 = \cos \beta_0$.  The first integral is
an integration over the spherical wind, which of course vanishes.
The second integral is over the cone, which scales with the factor $\eta$.

The separability means that the integration and radius and in angle can
be written as a product of factors, with

\begin{equation}
q = \frac{3}{16}\,\tau_0\,\sin^2 i_0\,\Gamma\,\Lambda,
\end{equation}

\noindent where

\begin{equation}
\Gamma = \int_1^\infty\,f(\tilde{r})\,D(\tilde{r})\,d\tilde{r},
\end{equation}

\noindent and

\begin{equation}
\Lambda = \int_{-1}^{+1}\,g(\mu)\,(1-3\mu^2)\,d\mu.
\end{equation}

\noindent For the problem at hand, and using a change of variable
with $\xi=\tilde{r}^{-1}$,
these functions become

\begin{equation}
\Gamma = \int_1^\infty\,\tilde{r}^{-2}\,w^{-1}\,D(\tilde{r})\,d\tilde{r} = \int_0^1\, \frac{D(\xi)}{w(\xi)}\,d\xi,
\end{equation}

\noindent and

\begin{equation}
\Lambda = \int_{\mu_0}^{+1}\,(1-3\mu^2)\,d\mu = -\mu_0\,(1-\mu_0)^2 \equiv \Lambda_0.
\end{equation}

\begin{figure}
\centering
\includegraphics[width=\hsize]{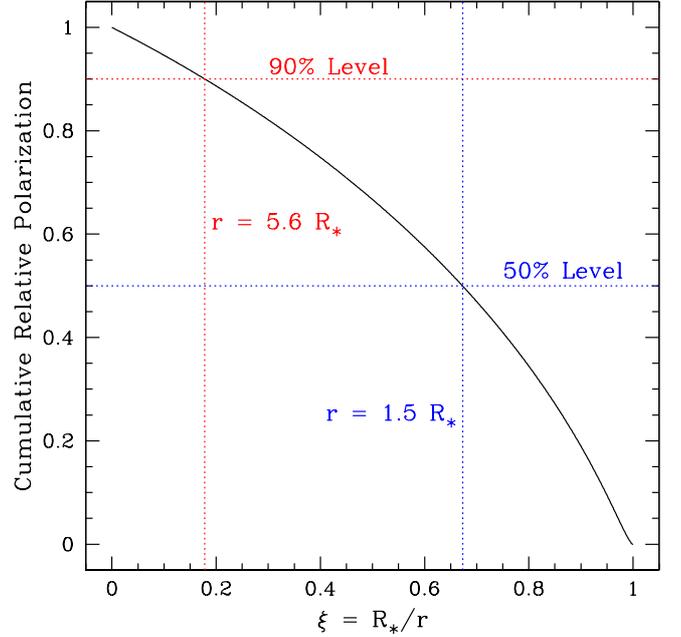}
\caption{Plot of the normalized cumulative polarization in terms
of distance from the star.  The figure applies when the polarization
calculation is seperable (see text).  In the case $\gamma=1$, 
references lines show that half of the polarization will be determined
within half a stellar radius above the atmosphere; 90\% will
be determined between one and 5.6 stellar radii.}
\label{fig:contrib}
\end{figure}

\noindent In relation to the radial integral factor, $\Gamma$, it is
useful to consider the sensitivity of the polarization to structure at
different radii in the wind.  Figure~\ref{fig:contrib} shows the value of
$\Gamma$ as integrated from $\xi$ to 1.  In this form the plot displays
$\Gamma$ as the cumulative polarization from the stellar surface out
to the radius $\tilde{r} = \xi^{-1}$ (for a given value of $\Lambda$),
hence the label for the vertical axis.  The plot is normalized to have
a unit value when the integral is carried out to infinite distance.
The reference lines show the locations where 50\% and 90\% of the
polarization are obtained.  In terms of the radial weighting, half of
the polarization is set between 1.0 and about 1.5 stellar radii; 90\%
is set between 1.0 and about 5.6 stellar radii.  The remaining 10\%
arises between $5.6R_\ast$ and beyond.  As a result, when discussing
the polarization due to a curved CIR, the bulk of the polarization
will be determined by the structure of the CIR at the inner few
stellar radii.

A word about convention is useful to mention at this point.  The observer system being used assumes the orientation of the cone as
projected in the sky lies along a north-south direction (i.e., along
a meridian).  For a cone with $\eta>0$, and noting that $\Lambda_0 <
0$, the net polarization would be negative. This simply indicates that the
electric vector of the net polarization is parallel to the east-west
line, which is orthogonal to the orientation of the cone projected onto
the sky.  If $\eta<0$, then $q>0$ because the polarization is dominated
by scatterers lying {\em \emph{outside}} the cone, and the polarization position
angle lies parallel to the cone.

Finally, that the cone orientation is north-south corresponds to
an orientation of the projected axis of symmetry on the sky as
$\psi=90^\circ$ as measured from west through north.  Obviously,
an arbitrary source could have an orientation on the sky, and
the net polarization would not change.  It is straightforward to
represent this polarization and position angle as follows.  Let $p_0 =
|q(\psi=90^\circ)|$.  Then for arbitrary orientation as perceived by an
observer, the normalized Stokes $q$ and $u$ parameters become

\begin{eqnarray}
q & = & p_0\,\cos(2\psi),~{\rm and} \\
u & = & p_0\,\sin(2\psi).
\end{eqnarray}

\noindent In developing a simplified model for the polarization from CIRs, we use
these constructions in what follows.

\subsection{CIRs in the equatorial plane}

A CIR amounts to a spiral-shaped zone of flow that threads a stellar
wind.  Its structure can indeed be complicated, as seen in the
hydrodynamic simulations of \citet{1996ApJ...462..469C} and 
\citet{2004A&A...423..693D}.  Our goal is to construct an approximation for the
structure of a CIR to rapidly explore model parameter space and
trends in the variable polarization with rotation phase.  To this
end, we approximate a cross-section of a CIR with a sphere as a
``spherical cap''.  This cap has circular cross-section and is
identical to the intersection of a sphere of radius $r$ with a cone.
Our construction for a CIR is simply these cap segments for a cone,
but where each successive segments in radius are azimuthally phase-lagged.  
In fact we require an equation of
motion for the centers of the segments.

Initially, we consider a CIR that is spiral in shape where the center
of the spiral lies in the equatorial plane.  The calculation for
the polarization can take an arbitrary function of
density into account {\em \emph{with radius}} inside the cone and even a varying
solid angle of the segments with radius.  As long as the CIR segments
are individually axisymmetric and scattering is optically thin, the
approach described in preceding sections can be employed.  However,
to reduce the number of free parameters, we assume that all segments
of the CIR have a constant solid angle, and we adopt the same
parameterization for the CIR density as for the cone of the preceding
section: the scaling with $\eta$ such that $\eta$ is a
constant throughout the CIR, and the $\gamma$ velocity exponent is
also the same in both the cone and the wind.

The question now is what to use for the equation of motion of
the segment centers.  What is the geometric form for the spiral?
A physically motivated approach could be based on wind-compression
theory \citep{1993ApJ...409..429B,1996ApJ...459..671I}.  That model is
for a rotating wind in which streamline flow is solved via kinematic
relations, namely a velocity law and conservation of angular momentum.
This model for rotating winds leads to density enhancements toward
equatorial latitudes at the expense of polar latitudes.  However,
wind-compression theory assumes a central force for the wind driving.
\citet{1996ApJ...472L.115O} have shown that this does not hold for
line-driven winds for massive stars.  Indeed, Owocki et al. find that including non-radial forces leads to density-enhanced polar flows.

For our purposes we adopt purely radial streamlines for wind flow
in the observer's frame.  This means that a parcel of fluid injected
into the wind from the star travels along a radial line, but since
the star rotates, subsequent parcels of gas also follow a radial
trajectory, but are successively displaced in azimuth.  Consequently,
the CIR amounts to a spiral pattern of flow that spins as a whole with
the rotation period of the star.  It is basically the same problem as
for perceiving a spiral pattern of water flow from a rotating hose.
The shape is found from the streamline flow for a CIR in the
rotating frame of the star and recognizing that the shape is equivalent
to the spiral pattern as seen by an observer.  (A similar approach was
used by \citet{1998ApJ...505..910I} to model the magnetic field topology
in the context of wind-compression theory.)

In the rotating frame of the star, we imagine a generalized ``spot''
of opening angle $\beta_0$ and solid angle $\Omega_0$ at the equator
of a star that rotates with period $P$ and angular velocity $\omega$.
This spot is the source of a density perturbation on the wind (i.e.,
the cause of $\eta \ne 0$).  In the rotating frame, the spot is fixed.
If we assume that the flow from the spot only has a radial component,
then in the rotating frame, a fluid element emerging from the spot
will progressively displace the spot azimuthally in proportion to $\omega t$,
where $t$ is the interval of time after the fluid parcel was injected
into the flow.

Generally, the fluid element should of course
emerge from the spot with an azimuthal speed of $\omega R_\ast$.
However, with the complexity of non-radial force considerations,
we simply ignore the details of the azimuthal component of
speed $v_\varphi$ that a fluid element would have.  Certainly,
a parametrization for $v_\varphi$ could be included.  In fact,
for $v_\varphi$ the radius-dependent azimuthal velocity in
the frame of the star, the equation of motion for the
center of the fluid element in the rotating frame 
of the star would be

\begin{equation}
\frac{d\varphi '}{dr} = \frac{1}{r}\,\left[\frac{v_\varphi(r) 
        - \omega r}{v_r}\right],
\end{equation}

\noindent where $\varphi '$ is the azimuth about the star in
the rotating frame as measured
from the spot (i.e., $\varphi ' =  0^\circ$ is the location of the
spot).  Our model for the spiral CIR assumes $v_\varphi(r) = 0$.

The solution to the preceding equation then becomes

\begin{equation}
\varphi ' = -\frac{R_\ast}{r_0}\,\int_\xi^1\,\frac{1}{\xi^2\,w(\xi)}\,d\xi,\end{equation}

\noindent where $r_0 = \vinf/\omega$ is a convenient parameterization
representing the ``winding radius'' of the CIR spiral.  Fast radial
flow means that the spiral is mostly a linear cone at radii close
to the star because the rotation is relatively slow, and so
$r_0/R_\ast$ is large, yielding $\varphi '$ is roughly constant for
$r < r_0$.  In contrast, fast rotation relative to the radial
flow speed indicates a rapid winding up of the spiral, which is
represented by a relatively low value of $r_0$.

In the observer's frame, the equation for the center of the spiral is

\begin{equation}
\varphi = \varphi_0 + \varphi ' (r) + \omega t,
\end{equation}

\noindent where $\varphi_0$ allows for the azimuthal location of the spot to be
defined in terms of an arbitrary reference.

For $\gamma=1$, the integral equation in $\xi$ is analytic.  The solution is

\begin{equation}
\varphi '(r=\xi^{-1}) = -\frac{R_\ast}{r_0}\,\left[ \frac{1-\xi}{\xi} +
        b\,\ln\,\left(\frac{w}{u\,w_0}\right) \right]
        \label{eq:eqmotion}
.\end{equation}

\begin{figure}
\centering
\includegraphics[width=\hsize]{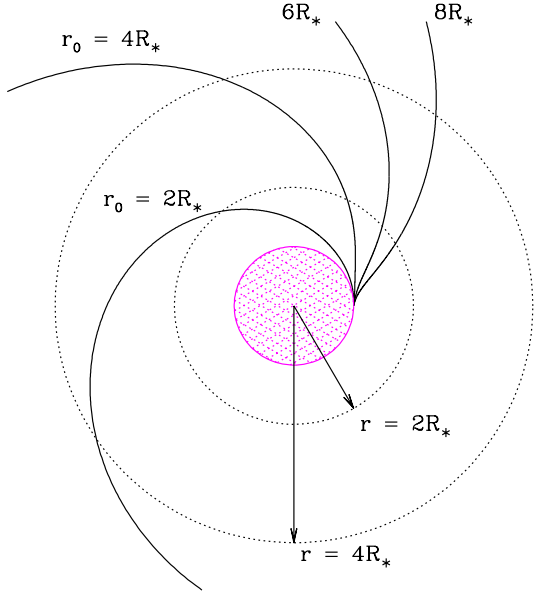}
\caption{Examples of the spiral shape for the CIRs as a function of
the winding radius, $r_0$.  The figure displays the {\em \emph{centroid}} of
four different CIRs with distance from the star.  The star is the
shaded magenta circle.  The CIRs are in the equatorial plane, shown
here as seen from above.  The two larger dotted circles are reference circles
at $2R_\ast$ and $4R_\ast$.  The different spirals are distinguished
by different winding radii as labeled.  Lower values of $r_0$
place the transition from a cone-shaped CIR to a spiral one closer to
the star; higher values of $r_0$ indicate that the CIR has small
curvature until achieving large distances from the star.}
\label{fig:r0}
\end{figure}

\noindent Analytic solutions can also be found for other integer
values of $\gamma$ (see the Appendix).  For our case of $\gamma=1$,
the influence of the winding radius, $r_0/R_\ast$, for the shape
of the CIR spiral is shown in Figure~\ref{fig:r0}.  Two reference
circles are shown for radii of $2R_\ast$ and $4R_\ast$.
The four curves are for the solution to $\varphi '$ for different
values of $r_0/R_\ast = 2, 4, 6,$ and 8, as labeled.  Lower values
of $r_0$ indicate relatively high equatorial rotation speeds
as compared to the wind terminal speed.  Curves with lower $r_0$ have
higher values of $\varphi '$ at a given radius.

To evaluate the polarization, a key result is to recognize the
separability of the spatial integrations for our adopted
model.  Our assumption is that
the cross-section for the CIR is the same as for a cone; therefore,
$\Lambda_0$ for the cone holds for the CIR.  The polarization for
a CIR reduces to evaluating $\Gamma$.  However, the solution for
$\Gamma$, which is an integration in radius, must take account of the
radius-dependent phase-lagging of segments as given by the solution of
equation~(\ref{eq:eqmotion}).

The phase-lagging must be related to the observer system of coordinates.
The result from Brown \& McLean reveals that the polarization from a
given conical segment scales with $\sin^2 i$.  For a CIR, each segment is
essentially at a different inclination and position angle.  In effect,
each segment has $i=i(\varphi(r,t))$ and $\psi = \psi(\varphi(r,t))$.
Using spherical trigonometry, one can show that

\begin{eqnarray}
\sin^2 i \cos 2\psi & = & \sin^2 \varphi - \cos^2 i_0\,\cos^2 \varphi,~{\rm and} \\
\sin^2 i \sin 2\psi & = & -\cos i_0\,\sin 2\varphi . 
\end{eqnarray}

\noindent Then the polarization for a CIR at a given a time in terms of the normalized
Stokes parameters $q$ and $u$ is given by

\begin{eqnarray}
q(t) & = & \frac{3}{16}\,\eta\,\tau_0\,\Lambda_0\,\Gamma_{\rm q},~{\rm and} \\
u(t) & = & \frac{3}{16}\,\eta\,\tau_0\,\Lambda_0\,\Gamma_{\rm u}, 
\end{eqnarray}

\noindent where

\begin{eqnarray}
\Gamma_{\rm q} & = & \int_0^1\,\frac{D(\xi)}{w(\xi)}\,\sin^2 i \, \cos 2\psi\,d\xi,~{\rm and} \\
\Gamma_{\rm u} & = & \int_0^1\,\frac{D(\xi)}{w(\xi)}\,\sin^2 i \, \sin 2\psi\,d\xi,\end{eqnarray}

\noindent with

\begin{eqnarray}
\sin^2 i \, \cos 2\psi & = & \sin^2 \varphi(t,\xi)-
        \cos^2 i_0\,\cos^2\varphi(t,\xi),~{\rm and} \\
\sin^2 i \, \sin2\psi & = & -\cos i_0\,\sin 2\varphi(t,\xi). 
\end{eqnarray}

\noindent The total polarization and the polarization position angle become

\begin{equation}
p = \frac{3}{16}\,\eta\,\tau_0\,\Lambda_0\,\sqrt{\Gamma_{\rm q}^2+\Gamma_{\rm u}^2},
\end{equation}

\noindent and

\begin{equation}
\tan 2\psi_{\rm p} = \frac{u}{q} = \frac{\Gamma_{\rm u}}{\Gamma_{\rm q}},
\end{equation}

\noindent where the last equality is specifically true for the many
assumptions that we have adopted, such as every cross-sectional
segment of the CIR having the same solid angle.  In the
optically thin limit, the polarization position angle is independent
of the wind optical depth, the value of $\eta$, and even the CIR
cross-section as encapsulated in $\Lambda_0$.  The only relevant
factor is the shape of the spiral as a function of radius as weighted
by the geometrical factors associated with Thomson scattering.

It is straightforward to include multiple CIR structures by
computing the $q$ and $u$ values for each one separately.  This holds
as long as each one is optically thin.  For $N$ CIRs at the equator,
each one having its own value of $\eta_{\rm i}$, $\varphi_{\rm 0,i}$,
and $\beta_{\rm i}$ for the i$^{th}$ CIR, 
the Stokes $q$ and $u$ parameters are linear
combinations with

\begin{eqnarray}
q_{\rm tot} & = & \sum_{\rm i=1}^N \, q_{\rm i},~{\rm and} \\
u_{\rm tot} & = & \sum_{\rm i=1}^N \, u_{\rm i}. 
\end{eqnarray}

\noindent However, it is useful to formulate the total mass loss of
the flow.  The angle-averaged electron optical depth of the entire
envelope becomes

\begin{equation}
\bar{\tau}_{\rm e} = \tau_{\rm w} \, \left[ 1 + \sum_{\rm i=1}^N \,
        \eta_{\rm i}\,\frac{\Omega_{\rm i}}{4\pi} \right]
        \label{eqsumm}
.\end{equation}

\noindent If one associates a certain $\dot{M}$ with the wind component,
then a total mass-loss rate can be determined from the preceding equation
after inclusion of CIRs.  

It is worth mentioning that CIRs  arise from redistribution of wind flow,
as contrasted to a model of sectors with enhanced or diminished mass loss
relative to an average spherical wind.  In redistribution one envisions
CIRs as perturbations of the wind density, and the summation term in
equation~(\ref{eqsumm}) must vanish.  In this framework
a ``CIR'' would actually consist of two CIR structures,  one with
positive $\eta$ and the other with negative $\eta$ so as to conserve the
mass-loss rate.  In other words redistribution of density implies that
$\bar{\tau}_{\rm e} = \tau_{\rm w}$, and our model would then suggest
pairs of CIRs with high ($\eta_{\rm h}$) and low ($\eta_{\rm l}$) densities,
such that

\begin{equation}
\eta_{\rm h}\Omega_{\rm h} + \eta_{\rm l}\Omega_{\rm l} = 0
\end{equation}

\noindent for every pair.  Model results to be presented in section~3
adopt a framework of high and low density streams instead
of a consideration of redistribution.  The point is that redistributive
approach could be modeled in a crude way with pairs of CIRs as
just described.

\subsection{CIRs from arbitrary latitudes}

Off the equator, the processes for formulating the CIR geometry and the
polarization is largely the same.  At the equator the CIR remains centered
in the equatorial plane.  At other latitudes the center of a CIR moves
along a conical surface under our scheme (since the flow is radial).  
The equation of motion for
$\varphi(t,r,\theta_0)$ for a stream originating at latitude $\theta_0$
is almost unchanged from the equatorial case.  With $\varpi = r \sin
\theta_0$, the differential equation for the centers of CIR segments is

\begin{equation}
\frac{d\varphi '}{d\varpi} = -\frac{\omega}{v_{\rm r}}.
\end{equation}

\noindent Given that for our approximation,
$\theta=\theta_0$ is fixed at all radii for a CIR
originating at $\theta_0$, the solution for $\varphi$ becomes

\begin{equation}
\varphi(t,\tilde{r},\theta_0) = \omega t + \varphi_0 - \frac{R_\ast}{r_0}\,
        \sin\theta_0\,\int_\xi^1\, \frac{d\xi}{\xi^2\,w}.
\end{equation}

\noindent The main difference is the appearance of the factor
$\sin\theta_0$.  The interpretation is that there is now an ``effective
winding radius'' with $r_{\rm eff} = r_0/\sin \theta_0$, and this becomes
larger for latitudes closer to the pole.  At the pole, $r_{\rm eff}
\rightarrow \infty$, and a CIR emerging from the pole would be a conical
stream.

A major difference from the equatorial case comes in the form of the
more complex spherical trigonometric relations that transform between
the star coordinate system and that of the observer.  The expressions
needed for $\Gamma_{\rm q}$ and $\Gamma_{\rm u}$ become

\begin{eqnarray}
\sin^2 i \, \cos 2\psi & = & \sin^2\theta_0\,\sin^2\varphi - \left(\cos\theta_0\,\sin i_0 \right. \nonumber \\
 & & \left. -\cos i_0\,\sin\theta_0\,\cos\varphi \right)^2,~{\rm and} \\
\sin^2 i \, \sin 2\psi & = & \sin i_0\,\sin 2\theta_0\,\sin\varphi \nonumber \\ 
 & & - \cos i_0\,\sin^2\theta_0\,\sin 2\varphi . 
\end{eqnarray}

\noindent With these relations in hand, polarization light curves can
be computed for a CIR emerging from an arbitrary location on the
star.  Indeed, using the approach of the preceding section, a
model with multiple CIRs can also be calculated.

The next section details an exploration of parameters as
a sampling of the kinds of variable polarizations that can
result for a number of select scenarios.

\begin{figure}
\centering
\includegraphics[width=\hsize]{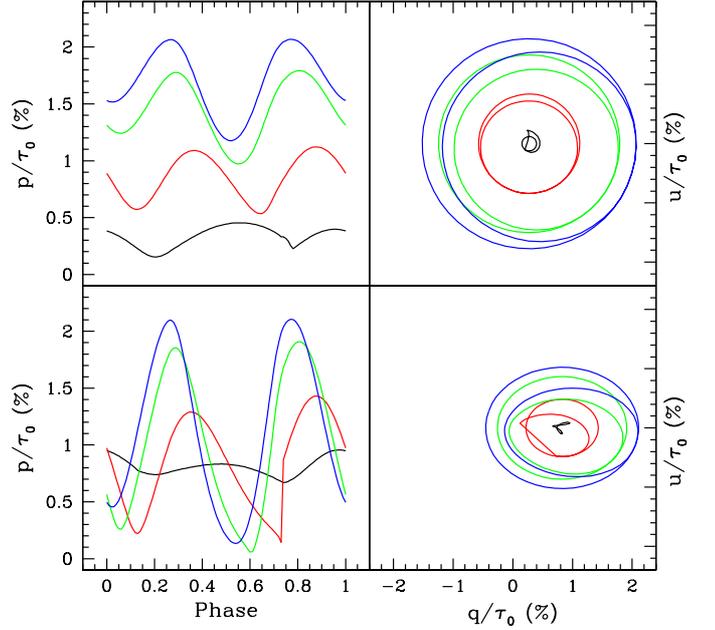}
\caption{
Periodic variations of polarization with stellar rotational phase
for CIRs with different winding radius values.  Left panels
are for polarization light curves with phase; right panels are
for variations in the $q$-$u$ plane.  Top panels are for a viewing
inclination of $i=30^\circ$ and lower for $i=60^\circ$.  Different curves
are for different winding radii, with values $r_0/R_\ast = 1$ (black),
3 (red), 9 (green), and 27 (blue).
}
\label{figr0}
\end{figure}

\section{Results}
\label{sec:results}

We have conducted a broad parameter study for the variable polarization
resulting from our model of CIRs.  Before discussing these results,
we wish to emphasize two points.  First, these models represent the
intrinsic source polarization.  Real measurements are contaminated
by an additional component arising from interstellar polarization.
An unresolved star that is spherically symmetric will generally
have a measurable polarization of the polarizing influence of the
interstellar gas and dust through which the starlight must travel
to reach Earth.  This influence depends on wavelength.  The effect
is frequently modeled or removed using a ``Serkowski Law'' 
\citep{1975ApJ...196..261S}.
Suppose the star has an intrinsic normalized polarization given by
$[q_\ast(\lambda),u_\ast(\lambda)]$ at the Earth. The interstellar
contribution can be modeled as adding an additional
$[q_{ISM}(\lambda),u_{ISM}(\lambda)]$.  The measured polarization
will then become $(q_\ast+q_{ISM},u_\ast+u_{ISM})$.  The ISM
contribution essentially produces a wavelength-dependent translation
of the intrinsic stellar polarization in the $q-u$ diagram.

Second, and most important for our purposes, the interstellar
contribution is constant.  Any variable polarization is therefore
necessarily intrinsic to the stellar source.  One cannot generally
say that the interstellar polarization is given by the observed
minimum $q$ and $u$ values.  In other words, even if the star did
not display variable polarization, it may still have a non-zero
intrinsic polarization.  Nonetheless, variable polarization is taken
as evidence for intrinsic polarization.  If the amplitude of that
polarization is larger than the minimum value, then one may
certainly conclude that the stellar polarization is large compared
to the interstellar component.

With these points in mind, what follows presents idealized
model results for $q_\ast(t)$ and $u_\ast(t)$ that ignore the ISM
component, as well as considerations of sources of random or systematic
(e.g., instrumental) errors present in the data.  Our assumption is that
our models would be applied to processed data after interstellar
and systematic effects have been corrected.

The parameter space for CIR simulations is large owing to our rather
general model.  Our model for the structure of CIRs is admittedly
parameteric.  We have strived to minimize the number of free
parameters in our approach.  Nonetheless, simulation variables
include  opening angle $\beta$, viewing inclination $i$, latitude
of the CIR $\vartheta$, azimuthal location $\varphi$, the winding
radius $r_0/R_\ast$, and the density enhancement factor $\eta$.  In
addition, simulations allow for an arbitrary number of CIR structures
$N$.

As a sampling of this enormous range of possible models, we focus
on four basic sets:  the effect of changing the winding
radius for one equatorial CIR, the effect of multiple CIRs all emerging from
the equator of the star, the effect of CIRs from different latitudes
as seen from different viewing inclinations, and the effect of
different numbers of randomly distributed CIRs as seen from a typical
viewing perspective.  Except for the first set of
models, the winding radius will be
fixed at $r_0/R_\ast = 5$.  This amounts to an equatorial rotation
speed that is 20\% of the wind terminal speed.  In addition, the
opening angle $\beta=15^\circ$ is fixed.  Thus at a given radius,
the CIR covers a solid angle of just under 2\% of the ``sky'' as
seen from the star.  Since our model is for optically thin
scattering, the polarization parameters $q$, $u$, and $p$ all scale
linearly with the optical depth $\tau_0$, so results will
be presented as normalized to $\tau_0$ as percentage values.

\subsection{Changing the winding radius}

Figure~\ref{figr0} shows model results for a single equatorial
CIR as the wind radius $r_0/R_\ast$ is varied.  Left panels
are polarization light curves; right are $q$-$u$ planes.
The top panels are for a viewing inclination of $i=30^\circ$,
whereas the bottom panels are for $i=90^\circ$.  The 
curves are for $r_0/R_\ast = 1, 3, 9,
and 27$.  Higher values for the winding radius imply that
CIRs start to curve at larger radii.

Not surprisingly, small winding radii tend to have low relative
polarization (i.e., $p/\tau_0$)
amplitudes.  This is because material is more spread
out in azimuth about the star, and so a greater range of scattering
angles are sampled, leading to polarimetric cancellation.  Higher
values of $r_0/R_\ast$ lead to higher relative polarization amplitudes.
In all cases, two peaks separated by half a rotational phase
are seen in the polarization light curve
because there is a single CIR structure.  In the $q$-$u$ plane,
these lead to a double loop.  The two troughs of the light curves
(and the two loops in the $q$-$u$ plane) are not identical because
of stellar occultation in conjunction with the CIR being
asymmetric.

\begin{figure}
\centering
\includegraphics[width=\hsize]{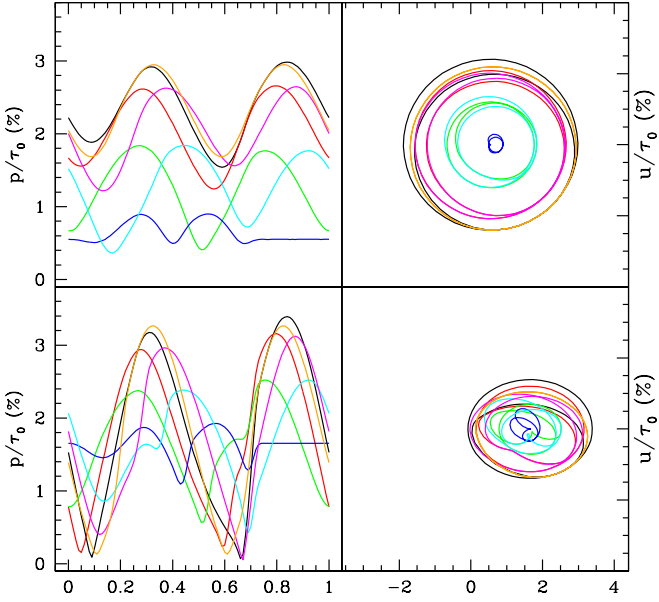}
\caption{
Periodic variable polarization for two
equatorial CIRs.  The format is the same as in Fig.~\ref{figr0} with
light curves to the left, $q$-$u$ plane variations to the right, a viewing
perspective of $i=30^\circ$ at the top, and $i=60^\circ$ along the bottom.
The two CIRs are shifted from each other by an angle $\Delta
\varphi$ of $0^\circ$ (black), $30^\circ$ (red), $60^\circ$ (green),
$90^\circ$ (dark blue), $120^\circ$ (light blue), $150^\circ$
(magenta), and $180^\circ$ (orange).
}
\label{figeq}
\end{figure}



\subsection{Equatorial CIRs}

We consider a model with two equatorial CIRs.  The first CIR has
$\eta=1$ and emerges from an azimuth of $\varphi = 0^\circ$.  A
second also has $\eta=1$ but emerges from $\varphi = 0^\circ$ to
$180^\circ$ in $30^\circ$ increments.  (Models from
$180^\circ$ to $360^\circ$ are degenerate with our set, within a
phase shift.) The two are co-added with $q=q_1+q_2$ and $u=u_1+u_2$.

Figure~\ref{figeq} displays polarimetric light curves for this set
of models.  The display is the same as in Figure~\ref{figr0}.  The
different colors for the different azimuthal offsets between the
pair of CIRs is indicated in the caption.  All of the light curves
are double peaked, but (a) the shape of the peaks, (b) the interval
between the peaks in phase, and (c) the amplitude of the peaks
depend on the relative positioning of the two CIRs in azimuth.
Of particular interest is that for the parameters shown, the
two peaks are closest when the CIRs emerge at $90^\circ$ from
one another around the equator.

\begin{figure}
\centering
\includegraphics[width=\hsize]{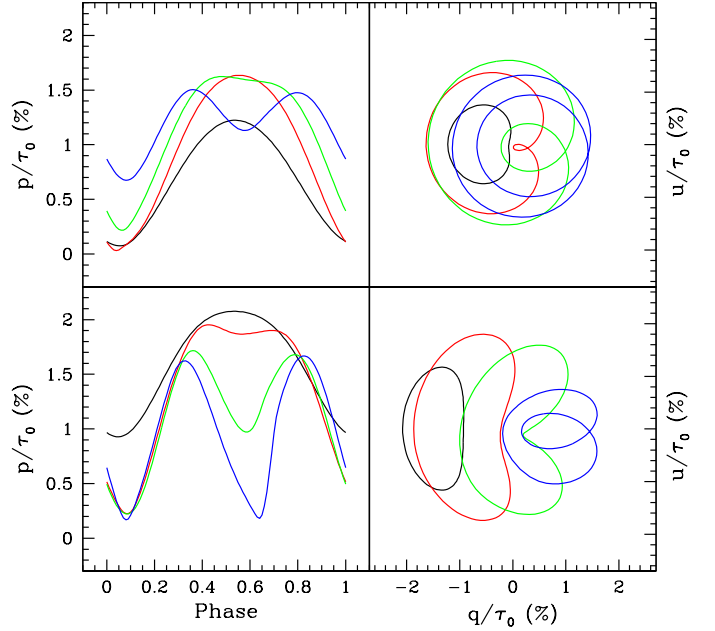}
\caption{
Periodic variations in polarization are shown here for CIRs that
emerge from different latitudes of the star.  The layout of the 
figure is the same as in Fig.~\ref{figeq}.  Here the colors are for
CIRs at latitudes of $\vartheta=20^\circ$ (black), $40^\circ$ (red), 
$60^\circ$ (green), and $80^\circ$ (blue).
}
\label{figlat}
\end{figure}

\subsection{CIRs at different latitudes}

In this set of models, a single CIR structure threads the wind, but
with the CIR emerging from different latitudes and viewed from
different inclinations.  Figure~\ref{figlat} shows results in
a format similar to Figure~\ref{figr0}.  Like that figure,
the upper panels are for $i=30^\circ$ and the lower for
$i=60^\circ$: left is for polarization light curves and right
for the $q$-$u$ plane.  
The different curves are for CIRs from different latitudes, with
$\vartheta = 20^\circ$, $40^\circ$, $60^\circ$, and $80^\circ$.

Changing the latitude has distinctive effects on the variable polarization
as compared to the previous two sections that considered only
equatorial CIRs.  Changing the latitude, all else being equal,
has consequences for whether the light curve is single or double
peaked.  In the $q$-$u$ plane, the effects are either a single
or a double loop.  The reason is that a CIR that is more closely located
near the pole tends to be more nearly stationary in projection
on the sky, despite the rotation of the star.  The limiting
case is a CIR at the pole, which is just a conical flow that maintains
a fixed projected position angle.

\begin{table}[b]
\caption{Model parameters for randomly distributed CIRs}
\label{tabrand}      
\centering          
\begin{tabular}{c c c c}        
\hline\hline                 
Case & $\eta$& $\varphi$ (degs) & $\cos\vartheta$ \\  
\hline                        
a & $-$0.64 & 220 & $+$0.18 \\
b & $+$1.83 & 250 & $-$0.89 \\
c & $+$0.20 & 219 & $-$0.22 \\
d & $+$0.48 & 290 & $+$0.98 \\
e & $+$0.63 &  14 & $+$0.22 \\
f & $-$1.94 & 326 & $-$0.63 \\
g & $+$0.52 & 154 & $-$0.74 \\
\hline                          
\end{tabular}
\end{table}

\subsection{Random distributions of CIRs}

The case that may perhaps most closely represent real stars is one
with multiple CIR structures emerging from different azimuth and
latitude locations.  In this example seven models have been
calculated with individual CIRs.  A random number generator was
used to place the individual CIRs around the star.  Values
of $\eta$ were randomly assigned in the range of $-1$ to $+4$.
Table~\ref{tabrand} summarizes the assigned parameters for the
seven cases.

\begin{figure}
\centering
\includegraphics[width=\hsize]{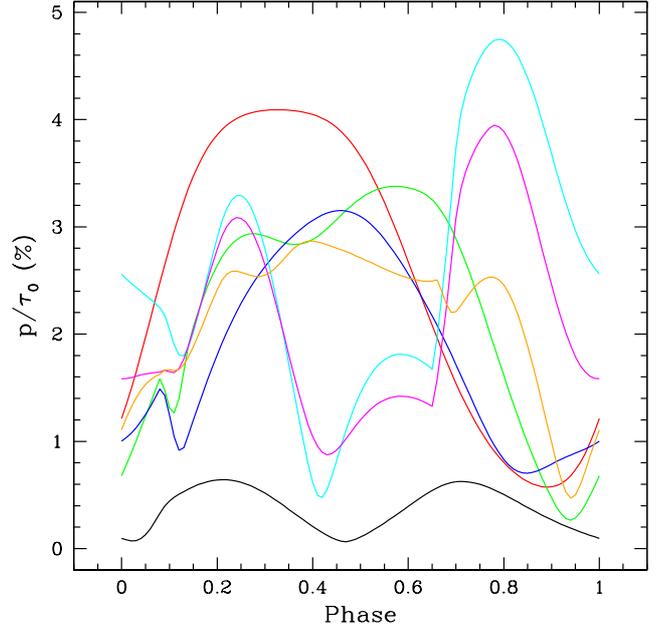}
\caption{
Periodic polarization light curves for a randomized distribution
of CIRs across the star.  Each curve represents the addition of
one more randomly placed CIR.  The curves correspond to a number
of CIRs of 1 (black), 2 (red), 3 (green), 4 (dark blue), 5 (light
blue), 6 (magenta), and 7 (orange).  Table~\ref{tabrand} lists
model parameters of the individual CIRs.
}
\label{figranda}
\end{figure}

\begin{figure}
\centering
\includegraphics[width=\hsize]{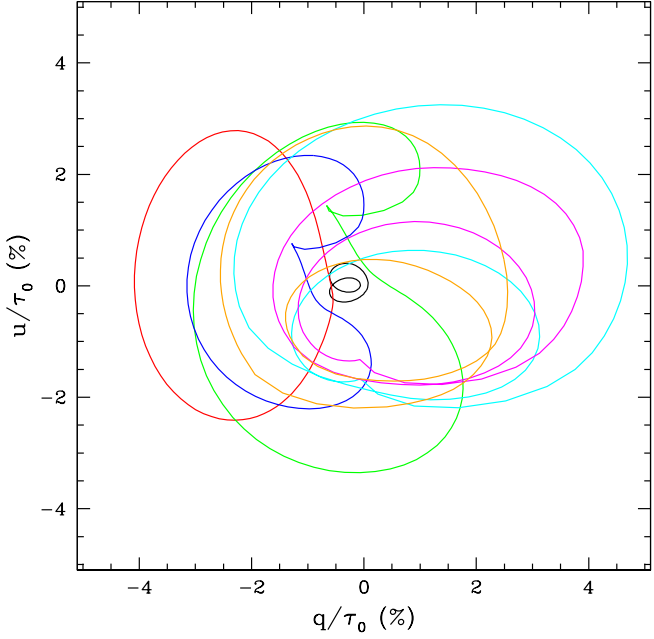}
\caption{
Polarization model results from Fig.~\ref{figranda} displayed in the $q$-$u$ plane.  The colors correspond to
the identifications indicated in Fig.~\ref{figranda}.
}
\label{figrandb}
\end{figure}

Figures~\ref{figranda} and \ref{figrandb} show polarization light curves 
and $q$-$u$ planes using
the following construction.  The viewing perspective is fixed
at $i=60^\circ$.  All the CIRs have $r_0/R_\ast=5$.  Then the results
are superimposed and normalized according to

\begin{equation}
q_{\rm net} = \frac{1}{n}\,\sum_{i=1}^{n}\, q_{\rm i}
\end{equation}

\noindent and

\begin{equation}
u_{\rm net} = \frac{1}{n}\,\sum_{i=1}^{n}\, u_{\rm i}
\end{equation}

\noindent where $n$ runs from 1 to 7.  Figures~\ref{figranda}
and \ref{figrandb} thus show results for one CIR, two CIRs, three CIRs, and so
on, up to all seven CIRs together.  This is an incremental approach for
seeing what happens as additional randomly placed CIR structures
are added to the model.

The normalization ensures that the curves are comparing 
the same things.  For example, if we were to co-add seven models of
exactly the same CIR, the resulting polarization would simply
be the same curve as one CIR, only with a polarization that is seven times
greater.  Normalization ensures that as more curves are added, the
curves are being compared on the same basis.

The result is a diverse set of possible outcomes.  The black curve
shows the case of just one CIR.  Adding additional CIRs moves through
the curves in color in the order of red, green, dark blue, light blue,
magenta, and finally orange.  A variety of shapes with more or fewer
peaks result.  Add to this the fact that for this example, all
CIRs have the same opening angle and that only one viewing
inclination case is being shown, it seems clear that a wide
variety of variable polarizations could result.

\section{Conclusion}
\label{sec:conc}

In this paper we have described a parametric representation of a
corotating interaction region (CIR) that threads an otherwise spherical wind.  There is no denying that we took
some liberties in presenting a simplified structure for a CIR:
circular cross-section, constant solid angle, uniform density within
the CIR, and a simple prescription for its spiral curvature.  However,
the advantages of the approach allow, for the first time, a broad
exploration of consequences for variable polarization signals.  The
goals were to adopt a flexible, semi-analytic description of a
CIR structure, basically motivated by hydrodynamic simulations, in
which to explore the range of polarimetric
behavior easily and rapidly.  These results may then be used in conjunction with
other methods, such as UV line variability \citep[for a recent example,
see][]{2012ApJ...759L..28P}, optical variability 
\citep[e.g.,][]{2010ApJ...716..929C}, and perhaps X-ray
variability \citep{2013ApJ...775...29I,2014MNRAS.441.2173M}.

An important result of our study is allowance for CIRs from latitudes
other than the equator.  The majority of previous works have focused on
equatorial CIRs \citep[one exception being][]{2004A&A...423..693D}.
The motivation is often one of simplicity.  Frequently, CIRs are
considered in relation to variable blueshifted absoprtion for UV line
data involving the intersection of the CIR with the absorption column
between the observer and the star.  Consideration of an equatorial CIR
makes the modeling easier.  In contrast, polarization is sensitive to the
fully three-dimensional structure of the asymmetric distribution of density
about the star.  (The trade-off in relation to line-profile studies is
the loss of velocity shift information.)  A major consequence of having
CIRs from different latitudes rather than just the equator is a far more
diverse range of polarimetric light curves and of behavior in the $q$-$u$
plane.  The potential-added complexity can be a strength when used in
conjunction with other diagnostics, such as line variability.

This contribution focused on the theoretical aspects of CIRs
and variable polarization.  There are several data sets to which
these results could be applied. In a follow-up paper, our model
will be employed to interpret variable polarization
for the Wolf-Rayet (WR) star EZ~CMa \citep{1989ApJ...343..426D}.  WR~stars are known to form
pseudo-photospheres, where optical depth unity occurs in the wind
itself \citep[e.g.,][]{1987ApJS...63..947H}.  Although our model is
for optically thin scattering, useful results can still be obtained
through adoption of the ``last scattering approximation'' \citep[e.g.,]
[]{1982SoPh...80..209S}.  The merit for such an approach is empirically
suggested by \citet{1992ApJ...387..347S} for EZ~CMa.  However,
the approach will require modifications to the prescription
for CIR curvature, which will be explained in the follow-up
paper.

\begin{acknowledgements}

The authors gratefully acknowledge helpful comments
from an anonymous referee.
RI wishes to acknowledge support for this research through a
grant from the National Science Foundation (AST-0807664).
NSL acknowledges financial support from the Natural
Sciences and Engineering Research (NSERC) of Canada.

\end{acknowledgements}

\begin{appendix}

\section{CIRs for the Case $\gamma=2$}
\label{sec:app}

\begin{figure}
\centering
\includegraphics[width=\hsize]{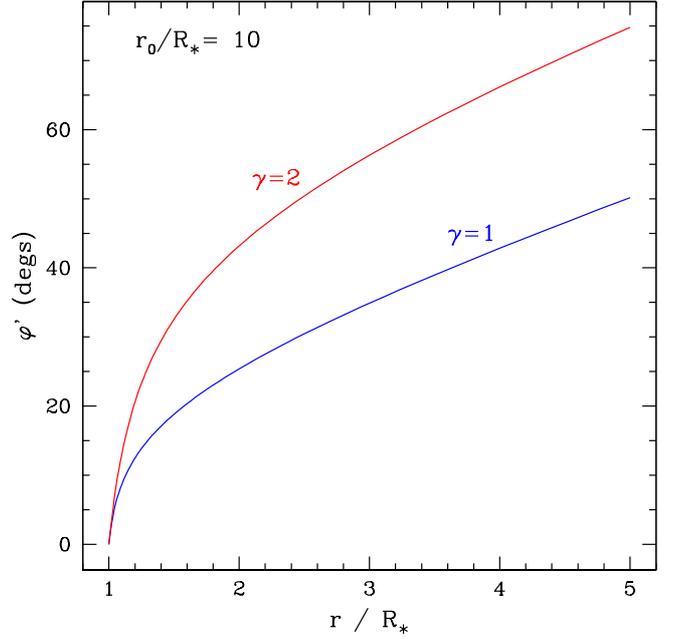}
\caption{Comparison of the azimuth location of the CIR centroid with
radius from the star.  Here blue is for $\gamma=1$, and red is for 
$\gamma=2$.  This example uses a wind radius of $r_0 = 10R_\ast$.  The
value of $\phi'$ is generally between 50\% to 100\% larger for $\gamma=2$
than for $\gamma=1$ indicating that a spiral CIR is considerably more
curved at a given radius for the higher $\gamma$ case.}
\label{figapp1}
\end{figure}

\begin{figure}
\centering
\includegraphics[width=\hsize]{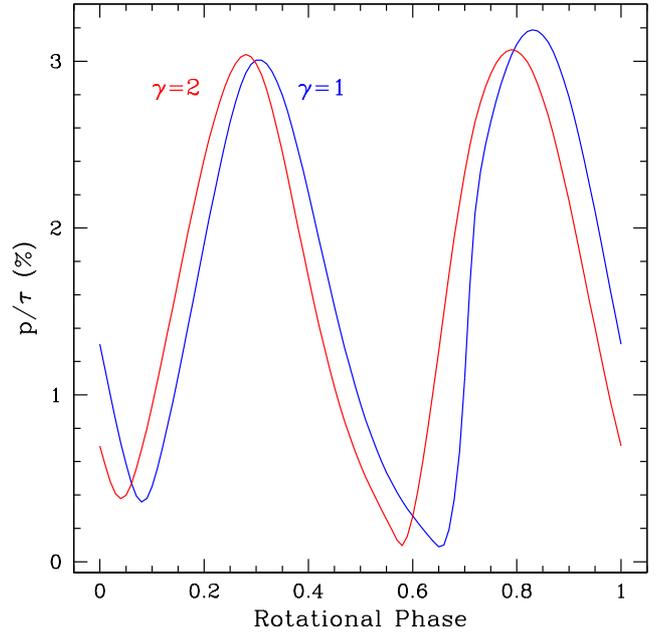}
\caption{Comparison of polarization light curves for the $\gamma=1$
and $\gamma=2$ cases shown in the previous figure.  The polarization
is plotted as normalized to the wind optical depth along a radial.
The case of $\gamma=1$ produces a relatively stronger polarization
signal (per unit optical depth) than in the case of $\gamma=2$.
The polarization extrema are slightly shifted in phase between
the two cases.}
\label{figapp2}
\end{figure}

Our results have focused exclusively on the case of $\gamma=1$
for the wind velocity law.  Of course, other values of $\gamma$
may be considered.  In general, the solution for the spiral shape
of the spiral will not be analytic but must be evaluated numerically.
However, analytic solutions for $\varphi(\xi)$ can be found for
integer values of $\gamma$.  Expected values of $\gamma$ for
early-type star winds range from $\gamma=0.5$ from the original
\citet{1975ApJ...195..157C} paper for line-driven winds to $\gamma\approx 0.8$
from \citet{1986A&A...164...86P} that augmented the initial results
of Castor et al. to the $\gamma=1$ that is typically used for
the inner wind of WR~stars \citep[e.g.,][]{1989A&A...210..236S} and then
to $\gamma \approx 3,$ which
has been suggested for some supergiant winds \citep[e.g.,][]{1995ApJ...452L..61P}.

As an example, the case of $\gamma=2$ is provided here as a comparison
case to $\gamma=1$.  In general, higher values of $\gamma$ tend to
increase the radial width over which the bulk of wind acceleration
takes place.  For example, with $w = (1-b/\tilde{r})^\gamma$, the radius
in the wind where the flow achieves half its terminal speed, $w=0.5$,
occurs at $\tilde{r}_{1/2} = b / (1-0.5^{1/\gamma})$, which increases
as $\gamma$ becomes larger.  As a case in point, $\tilde{r}_{1/2} = 2b$
for $\gamma=1$, but for $\gamma=2$, the value of $\tilde{r}_{1/2}$
increases to $3.4b$.
The net effect of this is that for our construction of the CIR spiral
shape, at a given value of $\omega$, a CIR tends to wind up at smaller
radius with increasing $\gamma$,  so the effective winding radius decreases
with larger $\gamma$.

We need to evaluate $\varphi '$ for $\gamma=2$.  We start with 
Eq.~(\ref{eq:varphi}) for a CIR in the equatorial plane, reproduced here:

\begin{equation}
\varphi ' = -\frac{R_\ast}{r_0}\,\int_\xi^1\,\frac{1}{\xi^2\,w(\xi)}\,d\xi.
\end{equation}

\noindent The wind velocity is $w = (1-b\xi)^2$.  Inserting $w$
into the above expression gives

\begin{equation}
\varphi ' = -\frac{R_\ast}{r_0}\,\int_\xi^1\,\frac{1}
        {\xi^2\,(1-b\xi)^2}\,d\xi.
\end{equation}

\noindent The solution for this expression is

\begin{equation}
\varphi ' = -\frac{R_\ast}{r_0}\,\left[\left(\frac{2b-1}{1-b}\right) 
        + \left(\frac{1-2b\xi}{\xi\,(1-b\xi)}\right) + 2b\,\ln
        \left(\frac{1-b\xi}{\xi\,(1-b)}\right) \right].
\end{equation}

\noindent Figure~\ref{figapp1} shows a comparison between these
functions.  There a value of $r_0/R_\ast =10$ is used.  
At most radii the value of
$\varphi '$ is about 50\% to 100\% higher for $\gamma=2$ than for
$\gamma=1$.  As compared to $\gamma=1$, the effective value of $r_0$
is about half as large when $\gamma=2$.

For the polarization one should also note that for a given optical
depth of the envelope, larger $\gamma$ essentially implies a relatively
higher density of scatterers at small radii as compared to winds with
lower $\gamma$ values.  Figure~\ref{figapp2} compares the polarization
light curves for the solutions displayed in
Figure~\ref{figapp1}:  equatorial CIRs, $r_0/R_\ast = 10$, $w_0 = 0.03$,
same opening angle, with $\eta=1$, and $i_0 = 60^\circ$ for both cases.
The polarization is plotted as the ratio $p/\tau$, where $\tau$ is the
optical depth of the wind.  As can be seen, for the
selected parameters that are relevant to fast winds, 
the polarization per unit
optical depth is similar for the two cases, but there is a slight phase shift
owing to the different degrees of CIR winding.
This result does not assume the same base density $n_0$ for
the two cases, but rather the same wind optical depth.  (For the same
$n_0$, the optical of a $\gamma=2$ wind is 1.6 times greater than for
$\gamma=1$.)

\end{appendix}

\bibliographystyle{aa}  
\bibliography{ignace}

\end{document}